\documentclass[submission,copyright,creativecommons]{eptcs}
\providecommand{\event}{GASCom 2024} 

\usepackage{iftex}

\usepackage{amsmath, amsthm, amssymb, amsfonts}
\usepackage{mathtools}
\DeclarePairedDelimiter{\ceil}{\lceil}{\rceil}

\usepackage{thmtools}
\usepackage{graphicx}
\usepackage{comment}
\usepackage{enumerate}
\usepackage[makeroom]{cancel}
\usepackage{float}
\usepackage{hyperref}
\usepackage[utf8]{inputenc}
\usepackage[english]{babel}
\usepackage{framed}
\usepackage[dvipsnames]{xcolor}
\usepackage{tikz}
\usetikzlibrary{shapes.geometric}
\usetikzlibrary{topaths}
\usetikzlibrary{decorations.pathreplacing,calligraphy}
\usepackage[T1]{fontenc}
\usepackage{caption}
\usepackage{subcaption}
\usepackage{forloop}
\usepackage{hyperref}
\usepackage[toc,page]{appendix}
\usepackage{dsfont}
\newcommand{\ts}{\textsuperscript}
\usepackage{stmaryrd}

\usepackage{todonotes}

\newcommand\pos[2]{Pos_#1(#2)}
\newcommand\cei{\ceil*{b/2}}
\newcommand\equi{\equiv}
\newcommand\True{\mathbf{T}}
\newcommand\False{\mathbf{F}}

\newcommand\grid[4]{
	\filldraw[step=1cm,gray,very thin,fill=white] (#3,#4) grid (#3+#1,#4+#2);
}

\newcommand\lcase[4]{
	\draw [draw=black, fill=#3] (#1, #2) -- (#1 +1, #2) -- (#1 +1, #2 +1) -- (#1, #2 +1) -- cycle;
	\node[] at (#1+0.5,#2+0.5){#4};
}

\newcommand\Ab{
	\mathcal{A}_b
}
\graphicspath{{images/}}

\newtheorem{theorem}{Theorem}

\declaretheoremstyle[name=Proposition,]{prosty}

\declaretheoremstyle[name=Principle,]{prcpsty}

\theoremstyle{definition}
\newtheorem{exmp}{Example}

\ifpdf
\usepackage{underscore}         
\usepackage[T1]{fontenc}        
\else
\usepackage{breakurl}           
\fi

\title{Counting Polyominoes in a Rectangle $b\times h$}
\author{Louis Marin
	\institute{UQAM LACIM\\ Montreal, Canada}
	\email{marin.louis@courrier.uqam.ca}
}
\def\titlerunning{Counting Polyominoes in a Rectangle $b\times h$}
\def\authorrunning{L. Marin}

\begin{document}
	\maketitle
	
	\bigskip
	\section{Introduction}
	The problem of enumerating polyominoes is known  to be a difficult problem since their introduction by Golomb \cite{Golomb} and so far,  only exhaustive generation computer programs are providing numerical answers.
	Nevertheless, restricted classes  of  these objects have been successfully enumerated \cite{bousquetmelou2008exactly,Zeilberger}.
	
	We focus on the problem of counting polyominoes inscribed in a rectangle of size $b \times h$.
	By \emph{inscribed}, we mean a polyomino that is included in a rectangle of size $b \times h$ and that has at least one cell touching each side of the rectangle.
	If we fix $b$ and increase $h$, the number of inscribed polyominoes satisfies a linear recurrence.
	The sequences for $b = 2,3,4$ are registered on the Online Encyclopedia of Integer Sequences (OEIS) (\textcolor{blue}{A034182} \cite{oeis}, \textcolor{blue}{A034184} \cite{oeis} and \textcolor{blue}{A034187} \cite{oeis}, respectively).
	The values for $5 \leq b \leq 12$ and $h = 24-b$ are also available (\textcolor{blue}{A292357} \cite{oeis}).
	
	The recurrence for $b = 2$ is known (\textcolor{blue}{A034182} \cite{oeis}) and can be proved using simple combinatorial argument.
	\begin{equation*}
		G_2 = \frac{2\,x^3+3\,x^2-2\,x+1}{\left(x-1\right)\,\left(x^2+2\,x-1\right)}
	\end{equation*}
	Where $G_b$ is the generating function for the number of polyominoes in a rectangle of size $b \times h$. 
	Recurrences for $b = 3, 4$ have been discovered empirically but, to the best of our knowledge, no proof seems to be available.
	
	In this extended abstract, we show how we obtained  the formulas for $b = 3,4,5,6$ and we design a method for obtaining the formulas for any $b$.
	To do so, we adapt methods described in previous works by Zeilberger \cite{Zeilberger} and by Bousquet-M\'elou and Brak \cite{bousquetmelou2008exactly} in order to build an automaton $\mathcal{A}_b$ recognizing exactly the polyominoes inscribed in a rectangle of fixed width $b$ and any height $h$.
	
	\section{Building $\mathcal{A}_b$}
	
	A polyomino $P$ is a set of edge-connected cells in the square lattice.
	If $P$ is inscribed in a rectangle of size $b \times h$, it can equivalently be described as a stack of $h$ rows of $b$ cells, where each cell is either selected or not and the selected cells are edge-connected.
	
	Each possible configuration in a row can be encoded by a unique word $u \in \{0,1\}^b, |u|_1>0$, a $0$ represents an empty cell, a $1$ represents a selected cell, and $|u|_1$ denotes the number of $1$'s occurring in the word $u$.
	A stack of height $h$ of such words encodes a unique polyomino \cite{bousquetmelou2008exactly}. In the next paragraphs, we use the expressions \emph{stack of words} and \emph{stack of rows} interchangeably.
	
	We wish to build an automaton $\mathcal{A}_b$ having the property that, given a stack of words, $\mathcal{A}_b$ accepts the stack if and only if it encodes a valid polyomino.
	Without loss of generality, we can assume that the automaton $\mathcal{A}_b$ operates by reading the stack of words from top to bottom.
	In order to accept only valid stacks, $\mathcal{A}_b$ needs to keep track of the connexity with each of its states. 
	If we extend the alphabet of the words $u$ to $\{0,1,...,\ceil*{b/2}\}$, we can label each cell of the row with a $0$ if the cell is empty and with the value $i \in \{1,...,\ceil*{b/2}\}$ if the cell belongs to the $i^{th}$ connected component of the polyomino, where the connected components are labeled with increasing values from left to right.
	
	We also want each state to encode whether the stack read so far has yet touched the left side or the right side of the rectangle.
	For this purpose, we introduce two boolean variables $l$ and $r$ that have value $\True$ if the leftmost (resp. rightmost) column has at least one selected cell, and $\False$ otherwise.
	
	From these conventions, we have that each state is of the form:
	\begin{equation*}
		(w,l,r) \quad \textrm{with} \quad w=w_1\cdots w_b \in \{0,1,...,\cei\}^b, \quad l,r \in \{\True, \: \False\}.
	\end{equation*}
	However, not every triplet of this form can suitably represent a valid polyomino row-configuration.
	We introduce the following additional conditions such that only the necessary triplets are kept:
	\begin{description}
		\item[\textit{Empty row}:] If $w = 0^b$, then $l = \False$ and $r = \False$.
		This row is forbidden in a polyomino but it is convenient to use this encoding as an initial state.
		
		\item[\textit{Inscription}:] If $w_1 \neq 0$, then $l = \True$, and if $w_b \neq 0$, then $r = \True$.
		This means that $P$ is adjacent to the left (resp. right) side  of the bounding rectangle.
		
		\item[\textit{Separation}:] If $w_i \neq 0$, then $w_{i-1}, w_{i+1} \in \{0,w_i\}$.
		In other words, if the $i^{th}$ cell of the current row is in $P$, each of its adjacent cells $w_{i-1}$ and $w_{i+1}$ is either empty or belongs to the same connected component, in virtue of the edge-connectedness requirement.
		
		\item[\textit{Non-crossing}:] Let $i,j,k,l \in \{1,2,...,\cei\}$ with $i < k < j < l$. Then the conditions $w_i = w_j$ and $w_k = w_l$ imply $w_i = w_k = w_j = w_l$.
		This condition comes from the fact that two distinct connected components cannot have crossed earlier in $P$.
	\end{description}
	
	We denote by  $\mathcal{T}_b$ the set of triplets $(w,l,r)$ that meet those conditions.
	
	Now, we still have the possibility that two distinct words in $\mathcal{T}_b$ represent the same row-configuration in $P$.
	For example the words $10201$ and $20102$ both represent the state where the first cell is connected above to the fifth cell, the third is in $P$ but disconnected from the first and fifth cells while the second and fourth cells are empty.
	
	To ensure injectivity, we introduce an equivalence relation.
	Let $w = w_1w_2...w_b \in A^b$ be a word on the alphabet $A$.
	For $a \in A$, we denote by  $\pos{a}{w}$  the set of all indices $i$ such that $w_i = a$.
	Then  $w$ and $w\prime$ are called \emph{equivalent}, and we write $w \equi w\prime$, if the following two conditions are verified:
	\begin{enumerate}[(i)]
		\item $\pos{0}{w} = \pos{0}{w\prime}$
		\item There exists $\sigma \in \mathfrak{S}_{\cei}$ such that $\pos{i}{w} = \pos{{\sigma(i)}}{w\prime}$, for $i \in \{1,2,...,\cei\}$.
	\end{enumerate}
	where $\mathfrak{S}_{\cei}$ is the symmetric group on $\cei$ elements.
	For each equivalence class of $\equi$, we choose as representative the minimum element with respect to the lexicographic order, denoted by $[w]$.
	
	We are now ready to define our automaton.
	More formally, let $\mathcal{A}_b = (\Sigma_b, Q_b, q_{0,b}, F_b, \delta_b)$, where
	\begin{enumerate}[(1)]
		\item{}$\Sigma_b = \{u \in \{0,1\}^b : |u|_1 > 0 \}$ is the set of possible words in a stack;
		\item{}
		$Q_b = \{([w],l,r) : [w] \in \mathcal{T}_b/\equi \}$ is the set of states;
		\item{}
		$q_{0,b} = (0^b,\False,\False)$ is the initial state;
		\item{}
		$F_b = \{([w],l,r) \in \mathcal{T}_b : w \in \{0,1\}^b, l = r = \True\}$ is the set of accepting states and
		\item{}
		$\delta_b: Q_b \times \Sigma_b \rightharpoonup Q_b$ is the transition function described in the next paragraphs.
	\end{enumerate}
	
	Observe that a state $([w],l,r) \in Q_b$ is also element of $F_b$ (i.e. is an accepting state) if and only if it has a single connected component (i.e. $w \in \{0,1\}^b$) and is inscribed in the rectangle (i.e. $l = r = \True$).
	The transition function $\delta_b$ is defined only on  the pairs $(([w],l,r),u)$, with $([w],l,r) \in Q_b, \: u \in \Sigma_b$,
	such that, for all $a \in \{1,2,\ldots,\cei\}, \: |\pos{a}{w}| > 0$ implies $\pos{a}{w} \cap \pos{1}{u} \neq \emptyset$.
	This ensures that, in the polyomino $P$, no connected component is ``lost''. 
	Since $u$ represents the configuration of the next row of the polyomino, $\pos{0}{u} = \pos{0}{[w^\prime]}$, and \[\delta_b(([w],l,r),u) = ([w^\prime], l^\prime, r^\prime),\]
	where $w^\prime$ is obtained by adding a row subject to the following constraints:
	
	\begin{description}
		\item[\textit{Vertical connexity}:] 
		From left to right, we read $u$ and write a new word $x$. 
		
		If $u_i = 0$ then $x_i = 0$. If $u_i = 1$ and $w_i \neq 0$, then $x_i = w_i$. If $u_i = 1$ and $w_i = 0$, then $x_i = N+1$ where $N = \max\{ w_1,w_2,...,w_b,x_1,x_2,...,x_{i-1}\}$.
		This step keeps the connexity between the cells in the next row and the cells in the current row that are directly above them. For the cells in the next row with no cells directly above them in the current row, we create new connected components for each of them.
		Observe that $x$ does not meet the \textit{separation condition}.
		
		\item[\textit{Horizontal connexity}:]
		From $x$, we create $w^\prime$. 
		
		We first obtain every factor of $x$ that is between $0$'s together with the largest prefix of $x$ that has no $0$'s and the largest suffix of $x$ that has no $0$'s. We call these factors the horizontally connected components of the new row. 
		We say that two horizontally connected components $c$ and $c^\prime$ are \textit{directly linked} if they share a letter and \textit{indirectly linked} if there exists a set $E = \{e_i \: | \: \text{$e_i$ is a horizontally connected component} \}$ such that $c=e_0$, $c^\prime = e_k$ and $e_i$ is \textit{directly linked} to $e_{i+1}$ for all $0 \leq i \leq n-1$ (i.e. there exists a chain of \textit{directly linked} connected components that links $c$ to $c^\prime$).
		The components $c$ and $c^\prime$ are said to be \textit{linked} if they are either \textit{directly linked} or \textit{indirectly linked}.
		
		The second step in the construction  of $w^\prime$ consists in collecting every \textit{linked} horizontally connected components into different sets. In other words, we take the transitive closure of the ``is linked to'' relation. Each set receives a letter in $\{1,2,...,\cei\}$. 
		The letters $\{1,2,...,\cei\}$ are assigned incrementally in increasing order to the set that contains the left-most horizontally connected component among the sets that have not yet been assigned a letter.
		
		Finally, we obtain $w^\prime$ from $x$ by looking at every horizontally connected component $c$ in $x$ and replacing each letter of $c$ by the letter assigned to the set to which $c$ belongs.
		
		\item[\textit{Adjacency}:] The values $l^\prime$ and $r^\prime$ are obtained by the computations $l^\prime = l \vee u_1$ and $r^\prime = r \vee u_b$.
	\end{description}
	\begin{exmp}Let
		\[
		w = 10203020104 \quad \text{and} \quad u = 10111011101.
		\]
		The first step yields the word $x = 10253026104$.
		The horizontally connected components are $1$, $253$, $261$ and $4$.
		The linked horizontally connected components are then joined together, forming the sets $\{1, 253, 261\}$ and $\{4\}$.
		The letters are associated with each set in order with the mapping
		\[1 \rightarrow \{1, 253, 261\}, \: 2 \rightarrow \{4\}.\]
		Finally, $w^\prime$ is obtained by replacing each letter in every horizontally connected components by the assigned letter, yielding $w^\prime = 10111011102$.
	\end{exmp}
	
	Observe that every step in the construction of $\Ab$ is automatic and $\delta_b$ is unambiguous. $\Ab$ is thus deterministic.
	
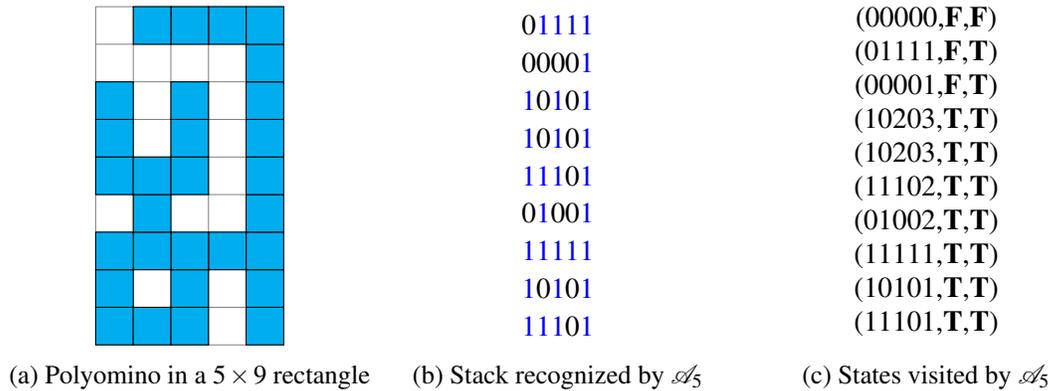
\begin{figure}[h]
		\centering
		\begin{subfigure}{0.3\textwidth}
			\centering
			\begin{tikzpicture}[scale = 0.5]
				\grid{5}{9}{0}{0}
				\lcase{0}{0}{cyan}{}
				\lcase{1}{0}{cyan}{}
				\lcase{2}{0}{cyan}{}
				\lcase{4}{0}{cyan}{}
				\lcase{4}{1}{cyan}{}
				\lcase{2}{1}{cyan}{}
				\lcase{0}{1}{cyan}{}
				\lcase{4}{2}{cyan}{}
				\lcase{3}{2}{cyan}{}
				\lcase{2}{2}{cyan}{}
				\lcase{1}{2}{cyan}{}
				\lcase{0}{2}{cyan}{}
				\lcase{4}{3}{cyan}{}
				\lcase{1}{3}{cyan}{}
				\lcase{4}{4}{cyan}{}
				\lcase{0}{4}{cyan}{}
				\lcase{1}{4}{cyan}{}
				\lcase{2}{4}{cyan}{}
				\lcase{4}{5}{cyan}{}
				\lcase{0}{5}{cyan}{}
				\lcase{2}{5}{cyan}{}
				\lcase{4}{6}{cyan}{}
				\lcase{0}{6}{cyan}{}
				\lcase{2}{6}{cyan}{}
				\lcase{4}{7}{cyan}{}
				\lcase{4}{8}{cyan}{}
				\lcase{3}{8}{cyan}{}
				\lcase{2}{8}{cyan}{}
				\lcase{1}{8}{cyan}{}
			\end{tikzpicture}
			\caption{Polyomino in a  $5 \times 9$ rectangle}
		\end{subfigure}
		\begin{subfigure}{0.3\textwidth}
			\centering
			\begin{tikzpicture}[scale = 0.5]
				\node[] at (0,0) {\textcolor{blue}{111}0\textcolor{blue}{1}};
				\node[] at (0,1) {\textcolor{blue}{1}0\textcolor{blue}{1}0\textcolor{blue}{1}};
				\node[] at (0,2) {\textcolor{blue}{11111}};
				\node[] at (0,3) {0\textcolor{blue}{1}00\textcolor{blue}{1}};
				\node[] at (0,4) {\textcolor{blue}{111}0\textcolor{blue}{1}};
				\node[] at (0,5) {\textcolor{blue}{1}0\textcolor{blue}{1}0\textcolor{blue}{1}};
				\node[] at (0,6) {\textcolor{blue}{1}0\textcolor{blue}{1}0\textcolor{blue}{1}};
				\node[] at (0,7) {0000\textcolor{blue}{1}};
				\node[] at (0,8) {0\textcolor{blue}{1111}};
			\end{tikzpicture}
			\caption{Stack recognized by  $\mathcal{A}_5$}
		\end{subfigure}
		\begin{subfigure}{0.3\textwidth}
			\centering {
				\begin{tikzpicture}[scale = 0.3]
					\node[] at (0,0) {(11101,$\True$,$\True$)};
					\node[] at (0,1.5) {(10101,$\True$,$\True$)};
					\node[] at (0,3) {(11111,$\True$,$\True$)};
					\node[] at (0,4.5) {(01002,$\True$,$\True$)};
					\node[] at (0,6) {(11102,$\True$,$\True$)};
					\node[] at (0,7.5) {(10203,$\True$,$\True$)};
					\node[] at (0,9) {(10203,$\True$,$\True$)};
					\node[] at (0,10.5) {(00001,$\False$,$\True$)};
					\node[] at (0,12) {(01111,$\False$,$\True$)};
					\node[] at (0,13.5) {(00000,$\False$,$\False$)};
			\end{tikzpicture}}
			\caption{States visited by $\mathcal{A}_5$ }
		\end{subfigure}
		\caption{Example of a word recognized by $\mathcal{A}_5$ and the associated polyomino}
		\label{Ex_A5}
	\end{figure}

	\begin{theorem}
		The set of stacks of size $h$ of words recognized by $\mathcal{A}_b$ is in bijection with the set of polyominoes inscribed in a rectangle of size $b \times h$. 
	\end{theorem}
	As an example,
	\autoref{Ex_A5} shows a stack of length 9 recognized by $\mathcal{A}_5$ and the associated polyomino.
	
\section{Generating the automaton $\mathcal{A}_b$.}
	
	We can manually compute the automata $\mathcal{A}_2$, $\mathcal{A}_3$ and, with a bit of courage, $\mathcal{A}_4$, $\mathcal{A}_5$ and $\mathcal{A}_6$.
	Using the Maple software, we were able to compute the generating functions for values of $b$ up to $6$.
	However, for $b \geq 7$, we need to turn to computers with more power, as the number of states grows rapidly.
	Thankfully, every step in the building of $\mathcal{A}_b$ can be automatized easily.
	
	With these methods, we computed the generating functions $G_b$ for $b = 3,4,5,6$. We also could compute $G'_b$ for $b = 3,4$ where $G'_b$ is the generating function for the number of polyominoes of area $n$ inscribed in a rectangle of size $b \times h$. The rational expression of those generating functions are of too large degree to include in this paper. Indeed, $G_3$ is of degree $9$, $G_4$ is of degree $20$, $G_5$ is of degree $49$ and $G_6$ is of degree $112$.
	
	The case $b=7$ was generated but the calculations to produce the generating function still need to be done.
	We are still working on the implementation to obtain the automata for greater values of $b$. The challenge comes from the fact that the systems grow exponentially in the number of states.
	Indeed, the number of states of $\mathcal{A}_b$ can be computed from the formula
	\begin{equation}
		\text{\# of states of } \mathcal{A}_b = 1 + \sum_{k=1}^{2^b-1} C_{f(k)}\cdot 2^{\llbracket k \equiv 0 \mod 2
			\rrbracket}\cdot 2^{\llbracket k < 2^{b-1} \rrbracket},
	\end{equation}
	where $C_m$ is the $m$\ts{th} Catalan number, $f(k)$ is the number of runs of $1$'s in the binary expansion of $k$ (\textcolor{blue}{A069010} \cite{oeis}) and $\llbracket p \rrbracket$ is the indicator function taking value $1$ if $p$ is true and $0$ otherwise.
	Surprisingly, the sequence of the number of states of $\mathcal{A}_b$ for increasing $b$ is not registered on OEIS.
	However, the sequence \textcolor{blue}{A140662} \cite{oeis}, also counting states in an automaton recognizing polyominoes, has been reported.
	\begin{figure}
		\begin{center}
			\begin{tabular}{ |l||l|l|l|l|l|l|l|l|l|l|l|l| }
				\hline
				$b$ & 0 & 1 & 2 & 3 & 4 & 5 & 6 & 7 & 8 & 9 & 10 & 11 \\
				\hline
				number of states in $\Ab$ & 1 & 2 & 6 & 16 & 40 & 99 & 247 & 625 & 1605 & 4178 & 11006 & 29292 \\
				\hline
			\end{tabular}
		\end{center}
		\caption{Number of states in $\Ab$ for $b = 0, \dots,11$}
	\end{figure}
	Avenues for immediate improvement include using a higher efficiency programming language and using a more suitable data type to store the states of the automaton.
	\bibliographystyle{eptcs}
	\bibliography{ref}
	\newpage

\end{document}